\documentstyle[12pt,epsf]{article}
\newcommand{\nc}{\newcommand}
\nc{\renc}{\renewcommand}
\renc{\baselinestretch}{1}
\nc{\com}[1]{\ \\ \ {\bf \# {#1}}\\ \ }
\nc{\bort}[1]{}
\nc{\be}[1]{\begin{equation} \mbox{$\label{#1}$}}
\nc{\bea}[1]{\begin{eqnarray} \mbox{$\label{#1}$}}
\nc{\Section}[2]{\section{\sc #2}\label{#1}}
\nc{\Subsection}[2]{\subsection{\sc #2}\label{#1}}
\nc{\Bibitem}[1]{\bibitem{#1}}
\nc{\Label}[1]{\label{#1}}
\nc{\eea}{\end{eqnarray}}
\nc{\ee}{\end{equation}}
\nc{\bdm}{\begin{displaymath}}
\nc{\edm}{\end{displaymath}}
\nc{\dpsty}{\displaystyle}
\nc{\bc}{\begin{center}}
\nc{\ec}{\end{center}}
\nc{\ba}{\begin{array}}
\nc{\ea}{\end{array}}
\nc{\bab}{\begin{abstract}}
\nc{\eab}{\end{abstract}}
\nc{\btab}{\begin{tabular}}
\nc{\etab}{\end{tabular}}
\nc{\bit}{\begin{itemize}}
\nc{\eit}{\end{itemize}}
\nc{\ben}{\begin{enumerate}}
\nc{\een}{\end{enumerate}}
\nc{\bfig}{\begin{figure}}
\nc{\efig}{\end{figure}}
\nc{\refc}[1]{\mbox{Ref.~\cite{#1}}}
\nc{\refs}[1]{\mbox{Refs.~\cite{#1}}}

\nc{\eqs}[2]{\mbox{Eqs.~(\ref{#1},\,\ref{#2})}}
\nc{\eq}[1]{\mbox{Eq.~(\ref{#1})}}

\nc{\figs}[2]{\mbox{Figs.~\ref{#1} and \ref{#2}}}
\nc{\fig}[1]{\mbox{Fig.~\ref{#1}}}

\nc{\figcap}[1]{\refstepcounter{figure}
        {\bf Figure \thefigure}: {\small\sl #1}}
\nc{\tabcap}[1]{\refstepcounter{table}
        {\bf Table \thetable}: {\small\sl #1}}

\nc{\tag}[1]{\label{#1} \marginpar{{\footnotesize #1}}}
\nc{\mtag}[1]{\label{#1} \mbox{\marginpar{{\footnotesize #1}}}}

\nc{\etal}{\mbox{\it et al. }}
\nc{\ie}{{\it i.e.}}
\nc{\eg}{{\it e.g.}}

\nc{\arreq}{&\!\!\!=\!\!\!&}
\nc{\arrmi}{&\!\!\!!-\!\!\!&}
\nc{\arrpl}{&\!\!\!+\!\!\!&}
\nc{\arrap}{&\!\!\!\approx\!\!\!&}
\nc{\non}{\nonumber}
\nc{\align}{\!\!\!\!\!\!\!\!&&}

\nc{\mat}[4]{{\left(\ba{cc} #1 & #2 \\ #3 & #4 \ea\right)}}

\def\simleq{\; \raise0.3ex\hbox{$<$\kern-0.75em
      \raise-1.1ex\hbox{$\sim$}}\; }
\def\simgeq{\; \raise0.3ex\hbox{$>$\kern-0.75em
      \raise-1.1ex\hbox{$\sim$}}\; }
\nc{\DOT}{\hspace{-0.08in}{\bf .}\hspace{0.1in}}
\nc{\Laada}{\hbox {$\sqcap$ \kern -1em $\sqcup$}}
\nc\loota{{\scriptstyle\sqcap\kern-0.55em\hbox{$\scriptstyle\sqcup$}}}
\nc\Loota{{\sqcap\kern-0.65em\hbox{$\sqcup$}}}
\nc\laada{\Loota}
\nc{\qed}{\hskip 3em \hbox{\BOX} \vskip 2ex}

\nc{\real}{{\rm I \! R}}
\nc{\Z}{{\sf Z \!\!\! Z}}
\nc{\complex}{{\rm C\!\!\! {\sf I}\,\,}}
\def\bigid{\leavevmode\hbox{\small1\kern-3.8pt\normalsize1}}
\def\id{\leavevmode\hbox{\small1\kern-3.3pt\normalsize1}}
\nc{\slask}{\!\!\!\!/}
\nc{\sslask}{\!\!\!/}
\nc{\bis}{{\prime\prime}}
\nc{\pa}{\partial}
\nc{\na}{\nabla}
\renc{\>}{\rangle}
\nc{\<}{\langle}
\nc{\goto}{\rightarrow}
\nc{\swap}{\leftrightarrow}

\nc{\EE}[1]{ \mbox{$\cdot10^{#1}$} }
\nc{\abs}[1]{\left|#1\right|}
\nc{\at}[2]{\left.#1\right|_{#2}}
\nc{\norm}[1]{\|#1\|}
\nc{\abscut}[2]{\abs{#1}_{\scriptscriptstyle#2}}
\nc{\vek}[1]{{\rm\bf #1}}
\nc{\integral}[2]{\int\limits_{#1}^{#2}}
\nc{\inv}[1]{\frac{1}{#1}}
\nc{\dd}[2]{{{\partial #1}\over{\partial #2}}}
\nc{\ddd}[2]{{{{\partial}^2 #1}\over{\partial {#2}^2}}}
\nc{\dddd}[3]{{{{\partial}^2 #1}\over
        {\partial #2 \partial #3}}}
\nc{\dder}[2]{{{d #1}\over{d #2}}}
\nc{\ddder}[2]{{{d^2 #1}\over{d {#2}^2}}}
\nc{\dddder}[3]{{d^2 #1}\over
        {d #2 d #3}}
\nc{\dx}[1]{d\,^{#1}x}
\nc{\dy}[1]{d\,^{#1}y}
\nc{\dz}[1]{d\,^{#1}z}
\nc{\dl}[1]{\frac{d\,^{#1}l}{(2\pi)^{#1}}}
\nc{\dk}[1]{\frac{d\,^{#1}k}{(2\pi)^{#1}}}
\nc{\dq}[1]{\frac{d\,^{#1}q}{(2\pi)^{#1}}}
\nc{\dbar}{d\!\!\!\stackrel{\stackrel{\!-}{}}{}\!\!\!}
\nc{\cc}{\mbox{$c.c.$ }}
\nc{\hc}{\mbox{$h.c.$ }}
\nc{\cf}{cf.\ }
\nc{\erfc}{{\rm erfc}}
\nc{\Tr}{{\rm Tr\,}}
\nc{\tr}{{\rm tr\,}}
\nc{\pol}{{\rm pol}}
\nc{\sign}{{\rm sign}}
\nc{\bfT}{{\bf T }}

\nc{\cA}{{\cal A}}
\nc{\cB}{{\cal B}}
\nc{\cD}{{\cal D}}
\nc{\cE}{{\cal E}}
\nc{\cF}{{\cal F}}
\nc{\cG}{{\cal G}}
\nc{\cH}{{\cal H}}
\nc{\cL}{{\cal L}}
\nc{\cM}{{\cal M}}
\nc{\cO}{{\cal O}}
\nc{\cT}{{\cal T}}
\nc{\rvac}[1]{|{\cal O}#1\rangle}
\nc{\lvac}[1]{\langle{\cal O}#1|}
\nc{\rvacb}[1]{|{\cal O}_\beta #1\rangle}
\nc{\lvacb}[1]{\langle{\cal O}_\beta #1 |}
\nc{\bb}{\bar{\beta}}
\nc{\ctH}{\tilde{\cal H}}
\nc{\chH}{\hat{\cal H}}
\nc{\1}{\aa}
\nc{\2}{\"{a}}
\nc{\3}{\"{o}}
\nc{\4}{\AA}
\nc{\5}{\"{A}}
\nc{\6}{\"{O}}
\nc{\al}{\alpha}
\nc{\Del}{\Delta}
\nc{\e}{\epsilon}
\nc{\eps}{\epsilon}
\nc{\lam}{\lambda}
\nc{\om}{\omega}
\nc{\Om}{\Omega}
\nc{\ve}{\varepsilon}
\nc{\mn}{{\mu\nu}}
\nc{\k}{\kappa}
\nc{\vp}{\varphi}

\nc{\pub}[4]{\Bibitem{#1}#2, {\sl ``#3''}, #4.}
\nc{\advp}[3]{{\it  Adv.\ in\ Phys.\ }{{\bf #1} {(#2)} {#3}}}
\nc{\annp}[3]{{\it  Ann.\ Phys.\ (N.Y.)\ }{{\bf #1} {(#2)} {#3}}}
\nc{\apl}[3]{{\it  Appl. Phys. Lett.\ }{{\bf #1} {(#2)} {#3}}}
\nc{\apj}[3]{{\it  Ap.\ J.\ }{{\bf #1} {(#2)} {#3}}}
\nc{\apjl}[3]{{\it  Ap.\ J.\ Lett.\ }{{\bf #1} {(#2)} {#3}}}
\nc{\app}[3]{{\it Astropart.\ Phys.\ }{{\bf #1} {(#2)} {#3}}}
\nc{\cjp}[3]{{\it  Can.\ J.\ Phys.\ }{{\bf #1} {(#2)} {#3}}}
\nc{\cmp}[3]{{\it  Comm.\ Math.\ Phys.\ }{{ \bf #1} {(#2)} {#3}}}
\nc{\cqg}[3]{{\it  Class.\ Quant.\ Grav.\ }{{\bf #1} {(#2)} {#3}}}
\nc{\epl}[3]{{\it  Europhys.\ Lett.\ }{{\bf #1} {(#2)} {#3}}}
\nc{\ijmp}[3]{{\it Int.\ J.\ Mod.\ Phys.\ }{{\bf #1} {(#2)} {#3}}}
\nc{\ijtp}[3]{{\it Int.\ J.\ Theor.\ Phys.\ }{{\bf #1} {(#2)} {#3}}}
\nc{\jmp}[3]{{\it  J.\ Math.\ Phys.\ }{{ \bf #1} {(#2)} {#3}}}
\nc{\jpa}[3]{{\it  J.\ Phys.\ A\ }{{\bf #1} {(#2)} {#3}}}
\nc{\jpc}[3]{{\it  J.\ Phys.\ C\ }{{\bf #1} {(#2)} {#3}}}
\nc{\jpg}[3]{{\it J.~Phys.~G:~Nucl.~Part.~Phys.~}{{\bf #1} {(#2)} {#3}}}
\nc{\jap}[3]{{\it J.\ Appl.\ Phys.\ }{{\bf #1} {(#2)} {#3}}}
\nc{\jpsj}[3]{{\it J.\ Phys.\ Soc.\ Japan\ }{{\bf #1} {(#2)} {#3}}}
\nc{\lmp}[3]{{\it Lett.\ Math.\ Phys.\ }{{\bf #1} {(#2)} {#3}}}
\nc{\lncim}[3]{{\it  Lett.\ Nuov.\ Cim.\ }{{\bf #1} {(#2)} {#3}}}
\nc{\mpl}[3]{{\it  Mod.\ Phys.\ Lett.\ }{{\bf #1} {(#2)} {#3}}}
\nc{\ncim}[3]{{\it  Nuov.\ Cim.\ }{{\bf #1} {(#2)} {#3}}}
\nc{\np}[3]{{\it  Nucl.\ Phys.\ }{{\bf #1} {(#2)} {#3}}}
\nc{\pr}[3]{{\it Phys.\ Rev.\ }{{\bf #1} {(#2)} {#3}}}
\nc{\pra}[3]{{\it  Phys.\ Rev.\ }{{\bf A#1} {(#2)} {#3}}}
\nc{\prb}[3]{{\it  Phys.\ Rev.\ }{{\bf B#1} {(#2)} {#3}}}
\nc{\prc}[3]{{\it  Phys.\ Rev.\ }{{\bf C#1} {(#2)} {#3}}}
\nc{\prd}[3]{{\it  Phys.\ Rev.\ }{{\bf D#1} {(#2)} {#3}}}
\nc{\prl}[3]{{\it Phys.\ Rev.\ Lett.\ }{{\bf #1} {(#2)} {#3}}}
\nc{\pl}[3]{{\it  Phys.\ Lett.\ }{{\bf #1} {(#2)} {#3}}}
\nc{\prep}[3]{{\it Phys\. Rep.\ }{{\bf #1} {(#2)} {#3}}}
\nc{\prsl}[3]{{\it Proc.\ R.\ Soc.\ London\ }{{\bf #1} {(#2)} {#3}}}
\nc{\ptp}[3]{{\it  Prog.\ Theor.\ Phys.\ }{{\bf #1} {(#2)} {#3}}}
\nc{\ptps}[3]{{\it  Prog\ Theor.\ Phys.\ suppl.\ }{{\bf #1} {(#2)} {#3}}}
\nc{\physa}[3]{{\it  Physica\ A\ }{{\bf #1} {(#2)} {#3}}}
\nc{\physb}[3]{{\it  Physica\ B\ }{{\bf #1} {(#2)} {#3}}}
\nc{\phys}[3]{{\it Physica\ }{{\bf #1} {(#2)} {#3}}}
\nc{\rmp}[3]{{\it  Rev.\ Mod.\ Phys.\ }{{\bf #1} {(#2)} {#3}}}
\nc{\rpp}[3]{{\it Rep.\ Prog.\ Phys.\ }{{\bf #1} {(#2)} {#3}}}
\nc{\sjnp}[3]{{\it Sov.\ J.\ Nucl.\ Phys.\ }{{\bf #1} {(#2)} {#3}}}
\nc{\spjetp}[3]{{\it Sov.\ Phys.\ JETP\ }{{\bf #1} {(#2)} {#3}}}
\nc{\yf}[3]{{\it Yad.\ Fiz.\ }{{\bf #1} {(#2)} {#3}}}
\nc{\zetp}[3]{{\it Zh.\ Eksp.\ Teor.\ Fiz.\ }{{\bf #1} {(#2)} {#3}}}
\nc{\zp}[3]{{\it Z.\ Phys.\ }{{\bf #1} {(#2)} {#3}}}
\nc{\zpc}[3]{{\it Z.\ Phys.\ C\ }{{\bf #1} {(#2)} {#3}}}
\nc{\ibid}[3]{{\sl ibid.\ }{{\bf #1} {#2} {#3}}}
\newcommand{\minus}{\!-\!}
\newcommand{\plus}{\!+\!}
\nc{\rf}[1]{(\ref{#1})}
\nc{\nn}{\nonumber \\*}
\nc{\Lbmeff}{\cL^{\beta,\mu}_{\rm eff}}
\nc{\Fmn}{F_{\mu\nu}}
\nc{\Psibar}{\overline{\Psi}}
\nc{\ati}{\tilde{a}}
\nc{\bt}{\tilde{b}}
\nc{\kt}{\tilde{k}}
\nc{\mt}{\tilde{m}}
\nc{\pti}{\tilde{\mbox{\boldmath $\Pi$}}}
\nc{\pv}{\mbox{\boldmath $\Pi$}}
\nc{\para}{\parallel}
\nc{\orto}{\perp}
\nc{\LL}{Landau level}
\nc{\LLL}{lowest Landau level}
\nc{\amm}{anomalous magnetic moment}
\nc{\dega}{\Delta E^{\beta}_{\gamma}}
\nc{\dee}{\Delta E^{\beta,\mu}_{e^+e^-}}
\nc{\dele}{\Delta E}

\nc{\dmbm}{\Delta m^{\beta,\mu}}
\nc{\mhz}{{\hat{\mu}_z}}
\begin{document}
\baselineskip 17pt
\begin{flushright}
  CERN-TH/95-274\\
  hep-ph/9510314
\end{flushright}
\begin{center}
{\Huge\bf   Dispersion Relations from the\\[3mm]
  Hard Thermal Loop Effective Action\\[3mm]
  in a Magnetic Field\\[5mm]}
\normalsize
\end{center}
\vspace*{10mm}
\bc
{\large Per Elmfors}
 \\[4mm]
{\sl TH-Division, CERN,
CH-1211 Geneva 23, Switzerland \\
Email: elmfors@cern.ch}
\ec

\vspace*{1mm}
\vfill
\bc
{\bf Abstract} \\
\ec
{\small
\begin{quotation}
\noindent
Dispersion relations for fermions at high temperature
and in a background magnetic field are calculated in two different
ways. First from a straightforward one-loop calculation where,
in the weak field limit, we find an expression closely related to
the standard dispersion relations in the absence of the magnetic field.
Secondly, we derive the dispersion relations directly from the
Hard Thermal Loop effective action, which allows for an exact
solution (i.e. to all orders in the external field),
up to the last numerical integrals.
\end{quotation}}
\vfill
  CERN-TH/95-274\\
  October 1995
\vspace{3mm}
\footnoterule
\noindent
{\small
Contribution to {\it The 4th International Workshop on
Thermal Field Theories and their Applications}, Dalian, China,
6 -- 12 August, 1995. }
\thispagestyle{empty}
\newpage
\setcounter{page}{1}
\section{Introduction}
\label{s:intr}
We know that thermal effects on dispersion relations are of
extreme importance at high temperature, where the whole concept
of  propagating particles is drastically
changed~\cite{Klimov82,Weldon8289}. Since these effects are large
and influence the mass shell conditions, it is also important
to include them consistently to all orders and not only as
first-order corrections. This is consistent to leading order in $(eT)^2$.
Both fermions and gauge bosons develop new branches and the zero-temperature
masses become less important.
New decay channels can open up because
of the effective thermal masses of the propagating excitations.
It is also interesting to study how strong external
fields influence the physics of these effective modes.
We have considered two ways of calculating the dispersion
relations for fermions at high temperature in a constant background
magnetic field. One method is to simply do the one-loop calculation
at finite temperature, using exact propagators in the external
field~\cite{ElmforsPS95}.
The other is to start from the already resummed effective action
for Hard Thermal Loops~\cite{BraatenP90,FrenkelTW90}
(HTLs). It is an effective action
that generates all HTLs at tree level and which contains the
external fields to all orders.
\section{One-loop self-energy}
\label{s:oneloop}
In Ref.~[3] the fermionic
dispersion relation at finite temperature
in a constant magnetic field was calculated using
the electron propagator in the Furry picture and
Schwinger's proper-time method.
In the high-temperature limit it turns out that
Schwinger's proper-time formulation~\cite{Schwinger51}
of the exact propagator is the easiest to
use for the loop calculation. The thermal propagator can be
constructed as~\cite{Tsai74}
\be{thprop}
        iS_{\rm vac}(p)-f_F(p_0)\biggl(iS_{\rm vac}(p)-
        iS_{\rm vac}^*(p)\biggr)\ ,
\ee
where for a magnetic field parallel to the $z$-direction%
\footnote{The sign convention here, that the particle has a
positive charge $e$, differs from Ref.~[3].}
\bea{Bprop}
        iS_{\rm vac}(p)&=& \int_0^\infty ds \frac{e^{ieBs\sigma_z}}{\cos eBs}
        \exp\left[is\left(p^2_\para-\frac{\tan eBs}{eBs}p^2_\orto-m^2
        +i\ve\right)\right]\nn
        &&\times\left\{\gamma p_\para-
        \frac{e^{-ieBs \sigma_z}}{\cos eBs}\gamma p_\orto
        +m\right\}\ .
\eea
After performing the loop integrals in
$\Sigma(x',x)=\<x'|\hat{\Sigma}|x\>$
it is possible to extract the gauge-invariant operator
$\hat{\Sigma}(p_0,p_z,\pv_\orto)$, where
$\pv_\orto=(p_x-eA_x,p_y-eA_y)$, in a gauge with zero
$z$-component. (We shall use the gauge $A_\mu=(0,0,-B\,x,0)$.)
Then we take the weak field limit, but keep $eB$
exactly wherever it is added linearly to the canonical momentum
like $\pv_\orto^2-eB\sigma_z$.
The reason being that $\pv_\orto^2-eB\sigma_z=\pv\slask_\orto^2$,
so $eB$ can be reabsorbed in $\pv\slask_\orto^2$,
which need not be small.
When the final expression for the self-energy is used in the effective
Dirac equation, we obtain
\bea{Deq}
        &&[\Pi\slask-m-\hat{\Sigma}(p_0,p_z,\pv_\orto)]\Psi=\nn[3mm]
        &&\left[s(p_0,\pv^2)\gamma_0p_0 -
        r(p_0,\pv^2)\gamma_zp_z-
        r(p_0,\pv^2-eB\sigma_z)\Pi\slask_\orto
        -m\right]\Psi=0~~,\quad
\eea
where $\pv^2=\pv_\orto^2+p_z^2$ and
\bea{sandr}
        s(p_0,\pv^2) &=& \left(1-\frac{\cM^2}{2p_0\abs{\pv}}
        \ln\abs{\frac{p_0+\abs{\pv}}{p_0-\abs{\pv}}}\right) \ ,
        \\[2mm]
        r(p_0,\pv^2) &=&\left(1+\frac{\cM^2}{\pv^2}
        \left(1-\frac{p_0}{2\abs{\pv}}
        \ln\abs{\frac{p_0+\abs{\pv}}{p_0-\abs{\pv}}}\right)\right)\ .
\eea
The temperature dependence enters only through the thermal mass
$\cM=e^2T^2/8$. It is {\em almost} possible to guess the expression
in \eq{Deq} from the standard expression for the HTL Dirac
equation~\cite{Klimov82,Weldon8289}.
The usual momentum $p_\mu$ should be replaced with the gauge-invariant
momentum $\Pi_\mu$, but there is an ambiguity in replacing $p^2$ by $\Pi^2$
or by $\Pi\slask\,\,\Pi\slask\,\,$. The correct way follows from the
calculations in Ref.~[3].

\bfig[ht]
   \epsfxsize=15cm
   \epsfbox{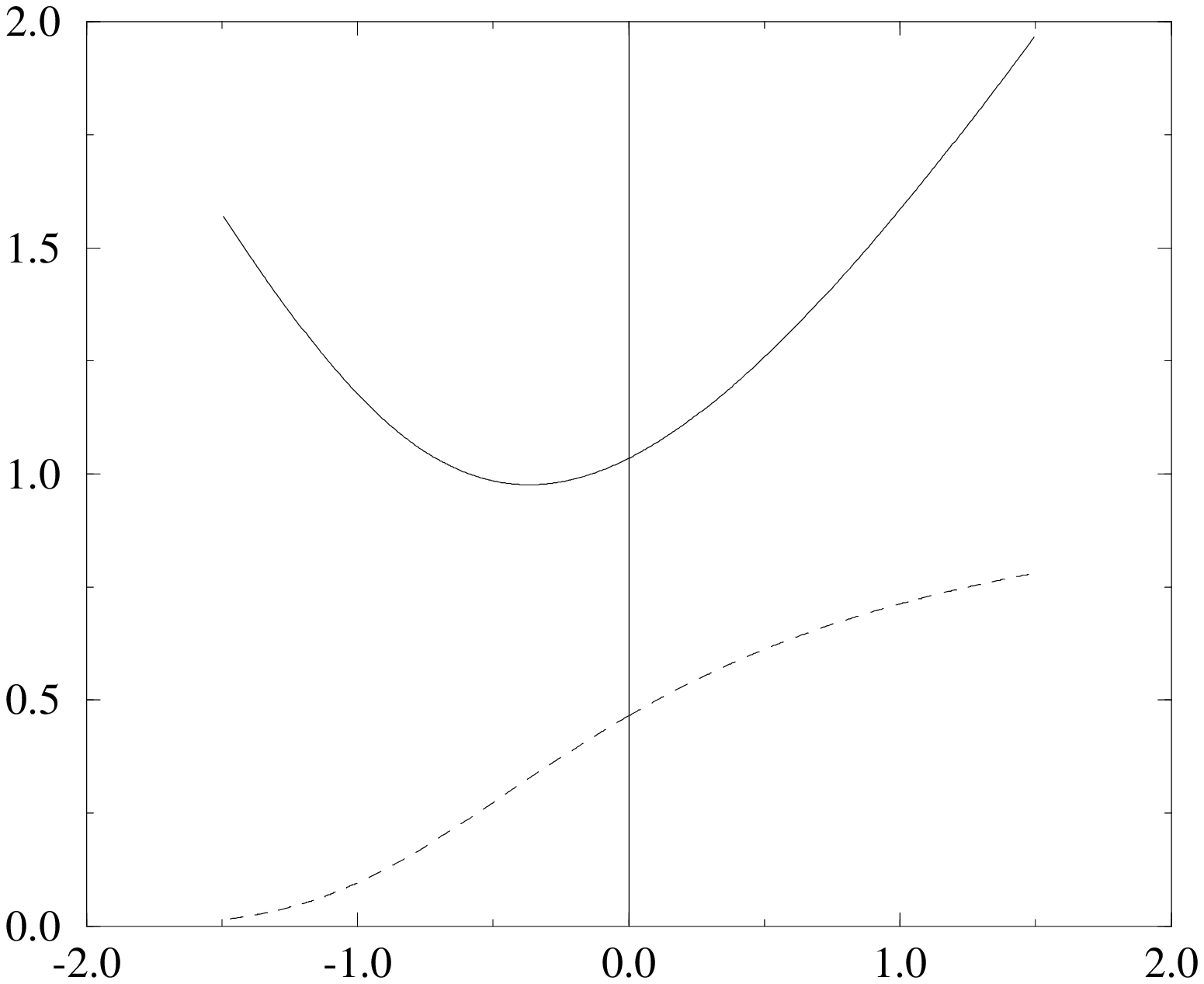}
   \setlength{\unitlength}{1.25mm}
\begin{picture}(0,0)(0,0)
   \put(0,0){}
   \put(62,12){\large $p_z$}
   \put(6,62){\large $E$}
   \put(45,30){\large $Z_h$}
   \put(75,47){\large $Z_p$}
   \put(48,63){\large $E_{Rh}$}
   \put(73,75){\large $E_{Rp}$}
   \put(40,85){\large $eB=0.2$}
   \put(40,80){\large $~~n=0$ \normalsize}
\end{picture}
   ~\\[-10mm]
   \figcap{Dispersion relation and spectral weight
   for the right-handed branch in the lowest Landau level.
   All dimensionful parameters are given in units of the thermal
   mass $\cM$.}
   \label{f:n0}
\efig
In the massless limit (i.e. zero vacuum mass) the dispersion relation
divides into a left- and a right-handed part. With a wave function
$\Psi=(L,R)^T$ in the chiral representation
the two dispersion relations become
\bea{LRdr}
        \align L:\quad (Es_n+p_zr_n)
        (Es_{n-1}-p_zr_{n-1})-2eBnr^2_{n-\inv{2}}=0~~, \\[2mm]
        \align R:\quad (Es_n-p_zr_n)
        (Es_{n-1}+p_zr_{n-1})-2eBnr^2_{n-\inv{2}}=0~~,
\eea
where $s_n=s(E,p_z^2+eB(2n+1))$ and similarly for $r_n$.
These relations are valid for all $n\geq 1$, but in the \LLL{} there is only
one non-zero component for each of $L$ and $R$,
so the dispersion relations reduce
to
\bea{LLLdr}
        \align L:\quad Es(E,p_z^2+eB)+p_zr(E,p_z^2+eB)=0\ , \\[2mm]
        \align R:\quad Es(E,p_z^2+eB)-p_zr(E,p_z^2+eB)=0\ .
\eea
The numerical solution in the \LLL{} is shown in \fig{f:n0} for the
right-handed particle. There are
eight different branches corresponding to the states
$(L/R)\times$ (particle/hole) $\times$ (positive/negative energy), for
each given value of $n\geq1$ and $p_y$. For $n=0$ there are only half
as many states, since only one spin orientation is possible in the \LLL.
Figure \ref{f:n0} only shows the right-handed branch,
and the left-handed
branch is obtained by reflection in $p_z$. In fact, from
\eqs{LRdr}{LLLdr} we find the combined symmetries
($L\leftrightarrow R$, $p_z\leftrightarrow-p_z$),
($L\leftrightarrow R, E\leftrightarrow -E$) and
($E\leftrightarrow -E, p_z\leftrightarrow -p_z$). In particular,
right-handed particles in the \LLL{} can only propagate along
the magnetic field (which points in the positive $z$-direction),
while the left-handed particles and the
right-handed holes propagate against the field.

The spectral weight presented in \fig{f:n0} is defined by
\be{Zdef}
        Z_i(p_z,n)^{-1}= \at{\frac{d}{d\om}}{\om=E_i(p_z,n)}
        \left(\Tr\left[(\cD(\om,p_z,n)\gamma_0)^{-1}\right]\right)^{-1}~~,
\ee
where $\cD$ is the $4\times4$ matrix in \eq{Deq} that
remains after acting on an off-shell \LL.
For a given chirality, $Z_i(p_z)$ is obviously very
asymmetric in $p_z$ since only particles propagate in
one direction and the holes in the other.
%
\section{HTL effective action in a background field}
\label{s:htl}
%
Instead of
computing the one-loop self-energy as in Section~\ref{s:oneloop},
we can use the fact that the HTL effective action already
contains the leading high-temperature contribution to all orders
in the gauge field. The equation of motion that follows from a
variation of the fermionic field should immediately give the
dispersion relation. The HTL effective action for
QED can be written as~\cite{Braaten93}:
\bea{HTLEA}
    \cL_{HTL}=\align-\inv{4}F^2+\frac{3}{4}\cM^2_\gamma
    F_{\mu\alpha}\left<\frac{u^\alpha u^\beta}
      {(\pa u)^2}\right> F_\beta^{~~\mu}\nn
      \align +\,\,\Psibar(i\pa\sslask-eA\sslask-m)\Psi
    -\cM^2\Psibar\gamma_\mu\left<\frac{u^\mu}{u\cdot\Pi}
      \right>\Psi~~,
\eea
where the average $\left<\cdot\right>$ is defined by
\be{ave}
    \left< f(u_0,\vec{u})\right>=\int \frac{d\Omega}{4\pi}
      f(1,\vec{u})~~,
\ee
where $\vec{u}$ is a spatial unit vector.
The equation of motion for $\Psi$ that follows is
\be{eqom}
    \left[\Pi\slask-m-\cM^2 \gamma_\mu
      \left<\frac{u^\mu}{u\cdot\Pi}\right>\right]\Psi=0~~.
\ee
Equation (\ref{eqom}) is a non-local and non-linear differential equation,
which is, in general, very difficult to deal with.
What makes this equation much less tractable than the
thermal Dirac equation in the absence of the $B$-field is that
the average over $\vec{u}$ is difficult to perform because
$[\Pi_\mu,\Pi_\nu]=-ieF_{\mu\nu}\neq 0$.
Since the spatial symmetries of the system are unaltered by the
thermal heat bath, we still expect  the eigenfunctions to have the
same spatial form as at zero temperature. In fact, after performing
the $u$-integral in \eq{eqom} the result can only be a function of the
invariants $\pv_\orto^2$, $p_0^2$ and $p_z$,
and the $\gamma$-structure has
to be proportional to $\gamma\pv_\orto$, $\gamma_0p_0$ and $\gamma_zp_z$.
We shall therefore compute the matrix elements
\be{matel}
    \<\Phi_{\kappa'}|\left<\frac{u^\mu}{u\cdot\Pi}\right>
    |\Phi_\kappa\>~~,
\ee
between the vacuum eigenstates
\bea{state}
    \<x|\Phi_{\kappa}\>&=&\exp[i(- p_0 t \plus  p_{y}y \plus  p_{z}z ) ]
    I_{n;p_y}(x) ~~,\\[2mm]
    I_{n;p_y}(x)&=& \left(\frac{eB}{\pi} \right)^{1/4} \exp \left[
    - \inv{2} eB \left( x \minus \frac{p_y}{eB} \right)^{2}\right]\nn
    &&\times\inv{\sqrt{n!}} H_n \left[\sqrt{2eB} \left(x-\frac{p_y}
    {eB} \right) \right] ~~,
\eea
where $\kappa=\{p_0,n,p_y,p_z\}$ and $H_n[x]$ are Hermite polynomials.
These states form a complete set of functions in four dimensions when the
energy is off shell.
In the chiral representation suitable spinors can be formed from
$\Phi_\kappa$ as
$\Psi_\kappa={\rm diag}[\Phi_\kappa,\Phi_{\kappa-1},\Phi_\kappa,%
\Phi_{\kappa-1}]\chi$ where $\chi$ is an undetermined
space-time-independent spinor.
The vacuum Dirac operator in \eq{eqom}
obviously gives
an eigenvalue when acting on $\Psi_\kappa$, but it is more
difficult to determine the action of the thermal part since $\Phi_\kappa$
cannot be an eigenfunction to $u\cdot\Pi$ for all $u$. One way to calculate
the matrix element in \eq{matel} is to find a basis such that
$v\cdot\Pi|v_p\>=v\cdot p|v_p\>$ and insert a unit operator
$\int d^4p |v_p\>\<v_p|$. This unit operator
is, of course, independent of $v$ after the $p$-integration, so
in particular we can choose $v=u$ and change the order of integrations.
After computing the matrix elements in \eq{matel} we find indeed
that they are diagonal in $\kappa$ for $u_0$ and $u_z$, and
have a mixing with the first subdiagonals for $u_x$ and $u_y$.
Define $\<u_{0,z,\pm}\>$ by
\bea{u}
     \<\Phi_{\kappa'}|\left<\frac{u_{0,z}}{u\cdot\Pi}\right>|\Phi_\kappa\>
     &=& (2\pi)^3\delta_{\kappa',\kappa}\<u_{0,z}\>_\kappa~~,\\[2mm]
     \<\Phi_{\kappa'}|\left<\frac{u_x\pm iu_y}{u\cdot\Pi}\right>|\Phi_\kappa\>
     &=& (2\pi)^3\delta_{\kappa',\kappa\mp1}\<u_\pm\>_\kappa~~,
\eea
and $\kappa\mp1=\{p_0,n\mp1,p_y,p_z\}$. These are exactly the components
that occur naturally when we include the $\gamma$-matrices
in the chiral representation.
The calculation of $\<u_{0,z,\pm}\>$ is a bit lengthy but straightforward,
and the result reads
\bea{u0}
    \<u_0\>_\kappa\arreq\inv{n!\sqrt{2\pi}}\int_{-\infty}^\infty
        ds\,H_n^2(s) e^{-s^2/2}
        \nn\align\times
        \left\{\frac{p_z}{2p^2}\ln\frac{p_0+p_z}{p_0-p_z}
          +\frac{Es\sqrt{2eB}}{2p^2\sqrt{p_0^2-p^2}}
          \arctan\frac{s\sqrt{2eB}}{2\sqrt{p_0^2-p^2}}\right\}~~,
\eea
\bea{uz}
    \<u_z\>_\kappa\arreq\inv{n!\sqrt{2\pi}}\int_{-\infty}^\infty
        ds\,H_n^2(s) e^{-s^2/2}
        \nn\align\times
        \left\{-\frac{p_z}{p^2}+
          \frac{p_0(2p_z^2-eBs^2)}{4p^4}\ln\frac{p_0+p_z}{p_0-p_z}
          \right.\nn\align\left.
          +\frac{p_z(2E^2-p^2)}{2p^4}
          \frac{s\sqrt{2eB}}{\sqrt{p_0^2-p^2}}
          \arctan\frac{s\sqrt{2eB}}{2\sqrt{p_0^2-p^2}}\right\}~~,
\eea
\bea{up}
    \<u_+\>_\kappa\arreq\frac{i}{\sqrt{2\pi n!(n-1)!}}\int_{-\infty}^\infty
        ds\,H_n(s)H_{n-1}(s) e^{-s^2/2}
        \nn\align\times
        \left\{\frac{s\sqrt{2eB}}{2p^2}
          -\frac{Esp_z\sqrt{2eB}}{2p^4}\ln\frac{p_0+p_z}{p_0-p_z}
          \right.\nn\align\left.
          +\frac{2p_z^2(p_0^2-p^2)-p_0^2 eB s^2}{2p^4\sqrt{p_0^2-p^2}}
          \arctan\frac{s\sqrt{2eB}}{2\sqrt{p_0^2-p^2}}\right\}~~,\\
    \<u_-\>_\kappa&=&-\<u_+\>_{\kappa+1}~~,
\label{um}
\eea
where $p^2=p_z^2+eBs^2/2$.
With these definitions the Dirac equation effectively reduces
to a $4\times4$ matrix in the spinor indices, since the other
quantum numbers have been diagonalized.
In the massless limit we take the determinant to find the
dispersion relations. They  factorize again as in
\eqs{LRdr}{LLLdr} and the equations for the right-handed
component are
\be{exdr}
\ba{rrl}
    n\geq 1:~&\Bigl(p_0-p_z-\cM^2
    (\<u_0\>_\kappa-\<u_z\>_\kappa)\Bigr)&\\[2mm]
    &\times \Bigl(p_0+p_z-\cM^2
    (\<u_0\>_{\kappa-1}+\<u_z\>_{\kappa-1})\Bigr)&\\[2mm]
    &-\left(\sqrt{2eBn}-i\cM^2\<u_+\>_\kappa\right)^2&=~0~~,\\[2mm]
    n=0:~& p_0-p_z-\cM^2
    (\<u_0\>_\kappa-\<u_z\>_\kappa)&=~0~~.
\ea
\ee
\bfig[ht]
   \epsfxsize=15cm
   \epsfbox{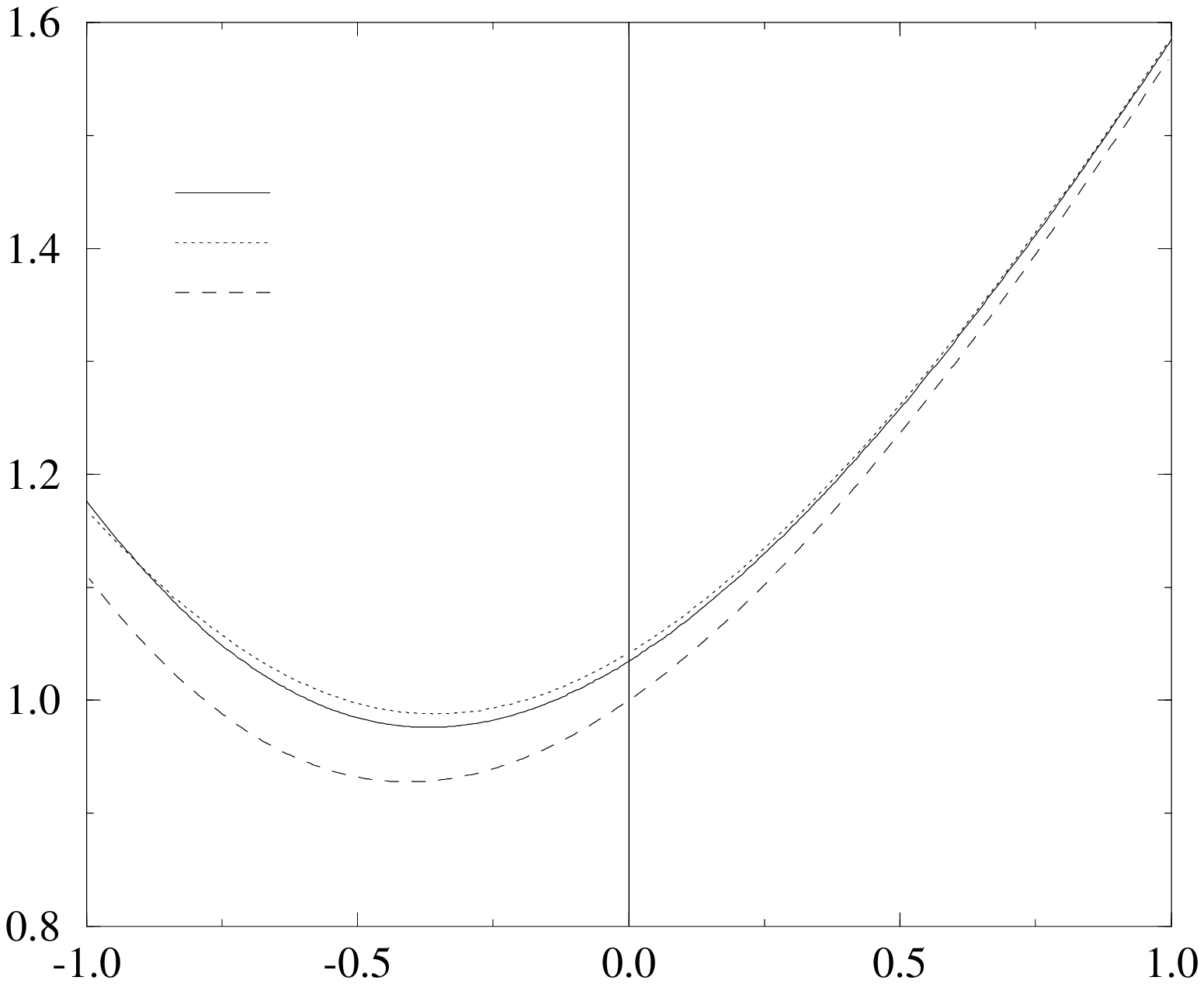}
   \setlength{\unitlength}{1.25mm}
\begin{picture}(0,0)(0,0)
   \put(0,0){}
   \put(62,12){\large $p_z$}
   \put(6,62){\large $E$}
   \put(39,80){ HTL}
   \put(39,76){ Weak field}
   \put(39,72){ $eB=0.0$}
   \put(70,85){\large $eB=0.2$}
   \put(70,80){\large $~~n=0$ \normalsize}
\end{picture}
   ~\\[-10mm]
   \figcap{Comparison of the dispersion relation from the HTL
     effective action and the weak field approximation.
   All dimensionful parameters are given in units of the thermal
   mass $\cM$.}
   \label{f:comp}
\efig
In general there are imaginary parts in the functions
$\<u_{0,z,\pm}\>_\kappa$, which make the spectral
functions more complicated.
The imaginary parts of Eqs.~(\ref{u0}) to (\ref{um}) are
determined by the analytic continuation $p_0\goto p_0+i\e$ for positive
$p_0$; this amounts to the replacement
\bea{repl}
    \align\inv{\sqrt{p_0^2-p^2}}\arctan\frac{s\sqrt{2eB}}{2\sqrt{p_0^2-p^2}}
    \nn
    \goto\align
    -\inv{2\sqrt{p^2-p_0^2}}\ln\frac{s\sqrt{2eB}+2\sqrt{p^2-p_0^2}}
    {s\sqrt{2eB}-2\sqrt{p^2-p_0^2}}
    -\frac{i\pi}{\sqrt{p^2-p_0^2}}~~,
\eea
for $p^2>p_0^2$.
It is anyway useful to solve \eq{exdr}, ignoring
the imaginary part, since the real part indicates where the
spectral functions are peaked, at least if the imaginary parts
are small enough. This can conveniently be done numerically
as all the integrals in Eqs.~(\ref{u0}) to (\ref{um}) are well convergent.
The result for the \LLL{} in a weak magnetic field ($eB=0.2\,\cM^2$) is
shown in \fig{f:comp}, and it agrees rather well with the
weak-field result from \eq{LLLdr}.
\bfig[ht]
   \epsfxsize=15cm
   \epsfbox{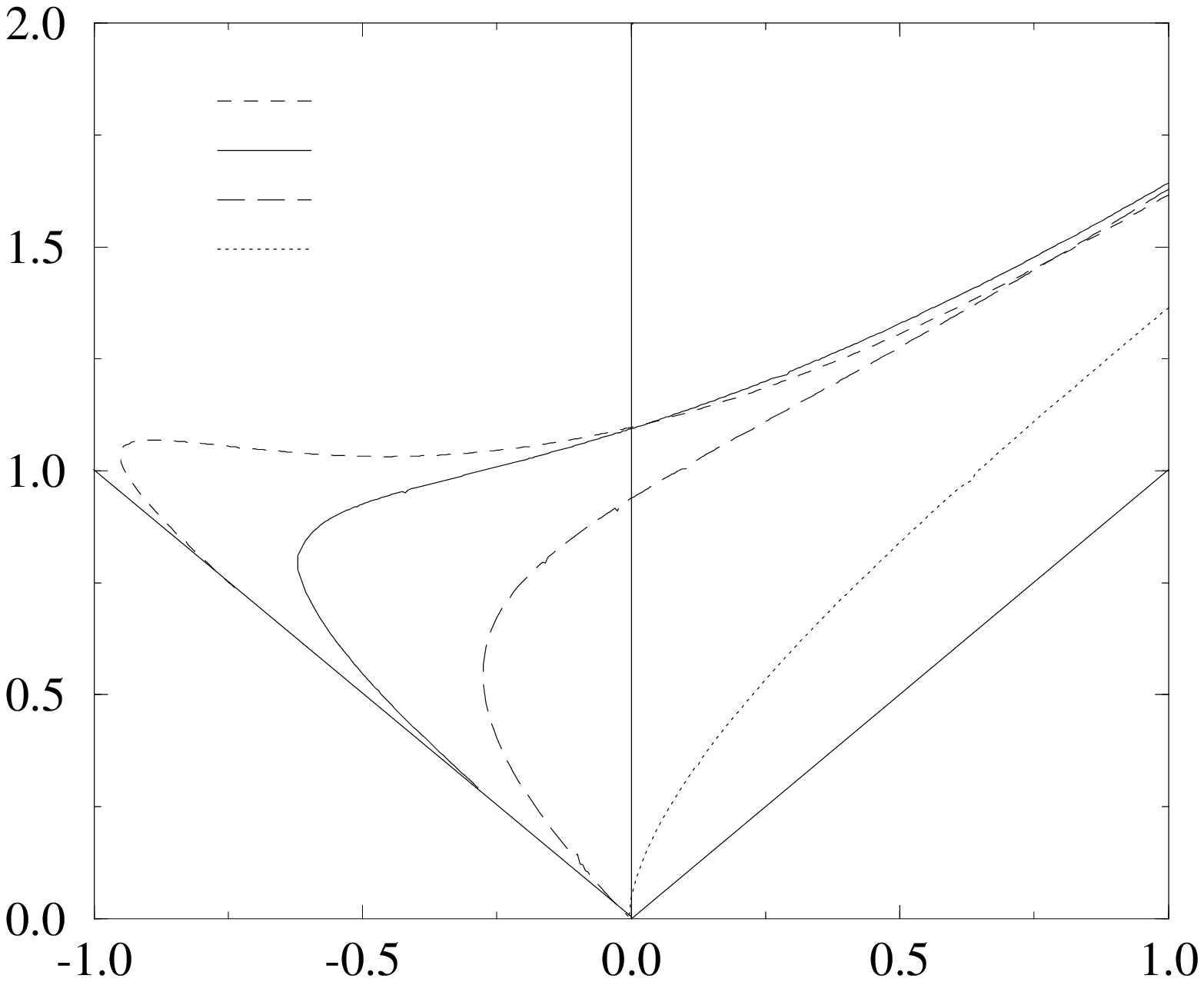}
   \setlength{\unitlength}{1.25mm}
\begin{picture}(0,0)(0,0)
   \put(0,0){}
   \put(62,12){\large $p_z$}
   \put(6,62){\large $E$}
   \put(42,88){ $eB=0.5$}
   \put(42,84){ $eB=1.0$}
   \put(42,80){ $eB=2.0$}
   \put(42,76){ $eB=10.0$}
\end{picture}
   ~\\[-10mm]
   \figcap{Dispersion relation for the right-handed branch in
     the lowest Landau level ($n=0$). As the $B$-field increases
     thermal effects become less important and
     the dispersion approaches the light cone, which is
     indicated by solid lines.
   All dimensionful parameters are given in units of the thermal
   mass $\cM$.}
   \label{f:n0.B1}
\efig

It is, of course, more interesting to see what happens at larger
field strength, which cannot be treated by \eq{LLLdr}. The dispersion
relations for  several field strengths are
shown in \fig{f:n0.B1}. Apart from the changes in the
particle branch there is a new branch coming
from the light cone, which eventually joins the hole branch
and disappears. This new branch can be understood mathematically,
from \eq{exdr}, by studying it close to the light cone. In the
absence of the $B$-field,
the hole branch exists because the logarithm in \eq{sandr}
becomes dominant close to the light cone, and the sign is
such that it allows for a positive energy solution to the part of that Dirac
equation which normally only gives the antiparticle solution.
In the present case, there is a compensation from the
new terms in \eq{exdr}, which change the behaviour close to the
light cone and a new branch can exist.
As the $B$-field increases the hole branch and its partner
become less extended, and the particle branch approaches a massless
mode. This is physically very reasonable since for very strong field
strengths  the thermal effects should disappear.

At this point it is worth discussing the approximations involved.
The HTL effective action is derived under the
assumption that the temperature is much
larger than the external momenta. Here, the momentum
is at least $\Pi\sim\sqrt{eB}$, which should be kept smaller than $T$.
But, if we consider the coupling constant $e$ to be very small
this approximation should be valid even for $eB\sim\cM^2$.
On the other hand, when $\Pi\gg T$ the vacuum part is dominant
and the dispersion relation is still approximately valid.
The other approximation was to neglect the imaginary part.
This approximation becomes worse for increasing
$eB$ and cannot be motivated for large $eB$.
A new massless excitation, such as the new hole partner,
could have important physical
consequences only if its spectral weight is non-negligible.
This remains to be studied and it can only be done correctly
using the full spectral function.
\section*{Acknowledgements}
I would like to thank David Persson and Bo-Sture Skagerstam for
collaboration on Ref.~[3], on which this contribution to a
large extent is based.
%

\end{document}